\documentstyle[12pt,epsf]{article}
\newcommand{\be}{\begin{equation}}
\newcommand{\ee}{\end{equation}}
\newcommand{\ba}{\begin{eqnarray}}
\newcommand{\ea}{\end{eqnarray}}
\newcommand{\bb}{\begin{thebibliography}}
\newcommand{\eb}{\end{thebibliography}}
\newcommand{\ci}[1]{\cite{#1}}
\newcommand{\bi}[1]{\bibitem{#1}}
\newcommand{\lab}[1]{\label{#1}}

\begin{document}
\phantom{.}
\vspace{2cm}
\begin{center}
{\bf {\large{Effects of Quark--Pomeron Coupling Structure in Diffractive
Deep Inelastic Scattering}}} \\

S.V.Goloskokov,
\footnote{Email:  goloskkv@thsun1.jinr.dubna.su}\\
Bogoliubov Laboratory of Theoretical Physics,\\ Joint Institute for
Nuclear Research,\\ Dubna 141980, Moscow region, Russia
\end{center}

\vspace{.7cm}
\begin{abstract}
We study the contribution of diffractive $Q \bar Q$ production
to the  $F_2^D$ proton structure function and the longitudinal double-spin
asymmetry in polarized deep--inelastic $lp$ scattering.  We show
the strong dependence of the $F_2^D$ structure function and the $A_{ll}$
asymmetry on the quark--pomeron coupling structure.

\end{abstract}
\newpage
\section{Introduction}

The diffractive events with a large rapidity gap between the final
proton $p'$ and the produced hadron system $X$ in deep inelastic
lepton--proton scattering

\be
e+p \to e'+p'+X                          \lab{de} \ee
have recently
 been studied  in several experiments \ci{gap1,gap2}.  The natural
explanation of these  events can be based on the hard photon--pomeron
interaction.  They can be interpreted as observation of the partonic
structure of the pomeron \cite{ing}.  Besides this effect, the
contribution where all the energy of the pomeron goes into the $Q \bar Q$
production \cite{lanj,coll} may be very important at small $x_p$ ($x_p$-
is  part of the momentum $p$ carried off by the pomeron). This region is
under investigation in the HERA energy domain.

The reaction (\ref{de}) can be described in terms
of the kinematic variables which are defined as follows:
\ba
Q^2=-q^2,\;t=(p-p')^2, \nonumber
\\  y=\frac{pq}{p_l p},\;x=\frac{Q^2}{2pq},\;
x_p=\frac{q(p-p')}{qp},\;\beta=\frac{x}{x_p},
\ea
where $p_l,p'_l$ and $p, p'$ are the initial and final lepton and proton
momenta, respectively, $q=p_l-p'_l$.

The diffractive structure function $F_2^{D(4)}$ is related to the
spin--average cross section of reaction (\ref{de})
\be
\frac{d^4\sigma}{dx dQ^2 dx_p dt}=
\frac{4 \pi \alpha^2}{x Q^4}[1-y+\frac{y^2}{2}] F_2^{D(4)}(x,Q^2,x_p,t).
\lab{f2}
\ee
Different models have been proposed to study the diffractive structure
function $F_2^{D}$ \cite{pom,fmodels}. As a rule, they are based on the
assumption about the factorization  at small $x_p$ of $F_2^{D}$ into the
pomeron flux factor $f(x_p,t)$ and the pomeron structure function
$F_2^{P}(\beta,Q^2,t)$

\be
 F_2^{D(4)}(x,Q^2,x_p,t)=f(x_p,t) F_2^{P}(\beta,Q^2,t).
 \ee
The partonic interpretation is usually used
for the pomeron structure function $F_2^{P}(\beta,Q^2,t)$ \ci{par}.

The function $f(x_p,t)$ at small $x_p$ behave as \ci{pom}
\be
f(x_p,t) \propto \frac{1}{x_p^{2\alpha_P(t)-1}},        \lab{flux}
\ee
where  $\alpha_P(t)$ is the pomeron trajectory
\be
\alpha_P(t)=\alpha_P(0)+\alpha' t,\;\;\;\; \alpha'=0.25(GeV)^{-2}.
\ee

The data on $F_2^{D}$ \ci{gap1,gap2} can be fitted by the single exponent
$x_p^{-N}$ where
\be
N \simeq  \begin{array}{l}
 1.2 \pm 0.1 \;\;\; {\rm for \;\;\; H1}\\
 1.3 \pm 0.1 \;\;\; {\rm for \;\;\; ZEUS}  \lab{n}
 \end{array}
 \ee
are independent of $\beta$ and $Q^2$. This behaviour is consistent with the
"soft" pomeron \ci{pom,soft} with
$ \alpha_P(0)= 1.1-1.15$.
 The hard perturbative pomeron \ci{bfkl} leads to $N \simeq 2$ that is in
 contradiction with HERA diffractive experiments.

Pomeron is a colour singlet state which is associated in QCD with
the two--gluon exchange \ci{low}. Gluons from the pomeron can interact
with a single quark in the loop, which leads to planar contributions.
There are nonplanar graphs in which gluons from the pomeron interact
with different quarks in the loop.  Such effects, as a rule, do not exceed
$10$ per cent as compared to the planar--diagram contributions (see e.g.
\cite{goljp}).  Moreover, it can be shown \ci{goltr} that the nonplanar
effects should be important at large momentum $k_{\perp}^2 \sim Q^2/
\beta$. The contribution of this region to the cross--section is rather
small.  As a result, the pomeron should prefer to interact with a single
quark in the loop. Thus, we can use the effective quark--pomeron vertex in
our calculations.

The diffractive scattering of polarized particles is proposed to be
studied at HERA and RHIC \ci{bu-now}. Then, the question of how large
the spin--flip component of the pomeron should be very important. In the
nonperturbative two-gluon exchange model \ci{la-na} and the BFKL model
\cite{bfkl} the pomeron couplings have a simple matrix structure (the
standard coupling in what follows):

\be
V^{\mu}_{hh I\hspace{-1.1mm}P}
=\beta_{hh I\hspace{-1.1mm}P}\; \gamma^{\mu}. \lab{pmu}
\ee
In this case,
the spin-flip effects are suppressed as a power of $s$.

The situation does change drastically when the large-distance loop
contributions are considered, which complicates the spin structure
of the pomeron coupling. These effects can be determined by the hadron wave
function for the pomeron-hadron couplings.
The spin-flip effects that do not vanish as $s \to \infty$ have been found
in some models of high-energy hadron scattering \ci{models,zpc}.  In the
model \ci{zpc} the pomeron--proton vertex has the form

\be
V_{ppI\hspace{-1.1mm}P}^{\mu}(p,r)=m p^{\mu} A(r)+ \gamma^{\mu} B(r),
\label{prver}
\ee
where $m$ is the proton mass. The coupling (\ref{prver}) leads to the
spin--flip in the pomeron exchange.

The spin structure of quark--pomeron coupling may not be so simple too.
It has been shown that the perturbative  gluon loop
corrections modify the form of the quark-pomeron coupling which looks like

\be
V_{qqI\hspace{-1.1mm}P}^{\mu}(k,r)= \beta_{qq I\hspace{-1.1mm}P}
[\gamma^{\mu} +2 M_Q k^{\mu} u_1+ 2 k^{\mu} /
\hspace{-2.3mm} k u_2 + i u_3 \epsilon^{\mu\alpha\beta\rho} k_\alpha
r_\beta \gamma_\rho \gamma_5+i M_Q u_4 \sigma^{\mu\alpha} r_\alpha],
\label{ver}
\ee
where $k$ is the quark momentum, $r$ is the momentum transfer and  $M_Q$
is the quark mass. So, in addition to the $\gamma_\mu$ term,
the new structures  appear from the loop diagrams.  The
functions $u_1(r)-u_4(r)$ are proportional to $\alpha_s$. It has been
shown \cite{gol4} that these functions can reach $20 - 30 \%$ of the
standard pomeron term $\sim \gamma_{\mu}$ for $|r^2| \simeq {\rm few}~
GeV^2$. Note that the modified  quark-pomeron coupling
(\ref{ver}) is drastically different from the standard one
(\ref{pmu}).  Really, the terms $u_1(r)-u_4(r)$ lead to the spin-flip at
the quark-pomeron vertex in contrast with the term $\gamma_\mu$.
We shall call the form (\ref{ver}) the spin-dependent pomeron coupling.
The phenomenological vertex $V_{qqI\hspace{-1.1mm}P}^{\mu}$
with the $\gamma_\mu$ and $u_1$ terms was proposed in \ci{klen}. The
modification of the standard pomeron vertex (\ref{pmu}) might be obtained
from the instanton contribution  \ci{inst}.

Thus, the pomeron couplings may have a complicated spin structure. As a
result, the spin asymmetries appear which have weak energy dependences as
$s \to \infty$.

In this paper, we study the effects of the spin--dependent quark--pomeron
coupling in diffractive deep inelastic scattering. We perform the
perturbative calculation of the diffractive $Q \bar Q$ production to the
process (\ref{de}) determined by Fig.1.  We investigate unpolarized
and polarized lepton-proton deep inelastic scattering that can be analyzed
in future polarized experiments at HERA.

In the second part of the paper, we study the diffractive contribution to
the $F_2$ structure function. In the third part, we calculate the
 distribution of spin--dependent cross sections and double--spin
longitudinal asymmetry over the transverse momentum of a produced jet. It
is shown that all these effects are sensitive to the quark--pomeron
coupling structure.

\section{$F_2^{D(3)}$
diffractive structure function}

 Let us study the diffractive jet production
 using expression (\ref{ver}) as the effective quark-pomeron coupling.
The term $\gamma^{\mu} B(r)$ in
the proton-pomeron coupling (\ref{prver}) gives the predominated
contribution to the cross sections for the spin-average and longitudinal
polarization of the proton.  Thus, in what follows we shall use the
standard form of the pomeron--proton coupling (\ref{pmu}).  The
spin--average cross section of the reaction (\ref{de}) can be written in
the form (\ref{f2}) where the diffractive structure function looks like

\ba
F_2^{D(4)}(x,Q^2,x_p,t) =\beta\frac{3 \beta_0^{4}
F(t)^{2}[9\sum_{i} e^2_i]}{1024\pi^{4} x_p} I(\beta,Q^2,x_p,t),
\lab{f2d}\\
I(\beta,Q^2,x_p,t)=\int_{k^2_0}^{Q^2/4\beta} \frac{d k_\perp^2
N(\beta,k_\perp^2,x_p,t)}
{\sqrt{1-4k_\perp^2\beta/Q^2}(k_\perp^2+M_Q^2)^2}.
\lab{int}
\ea
Here  $M_Q$ is the quark mass,
$\beta_0$ is the quark--pomeron coupling, $F(t)$
is the pomeron-proton form factor, $e_i$ are the quark charges.
The integral $I(\beta,Q^2,x_p,t)$ represents the contributions of the box
diagram of Fig.1 and the corresponding crossed graph. The function $N$ is
determined by the trace over the quark loop. It can be written for $x_p
=0$ in the form

\be
 N(\beta,k_\perp^2,t)= N^s(\beta,k_\perp^2,t)+
 \delta N(\beta,k_\perp^2,t).   \lab{nd}
 \ee
Here $N^s$ is the contribution of the standard pomeron vertex
(\ref{pmu}) and $\delta N$ contains the contribution of the $u_1(r)-u_4(r)$
 terms from (\ref{ver}). The traces have been calculated by using the
 programme REDUCE. For the term $N^s$ in the case of light quarks in the
loop we find
\be
N^s(\beta,k_\perp^2,t)=32 [2(1-\beta) k_\perp^2-\beta |t|]|t|. \lab{ns}
\ee
 The calculation of the $\delta N$ term is  difficult.
We have found it in the $\beta \to 0$ limit. For the massless
quarks only the $u_3$ terms contribute to $\delta N$:
\be
\delta N(k_\perp^2,t) = 32 k_\perp^2 |t| [(k_\perp^4+4
k_\perp^2 |t|+|t|^2) u_3- 4 k_\perp^2-2 |t|] u_3.  \lab{deld}
\ee
Note that $\delta N$ is positive because $u_3 \le 0$. Higher twist terms
of an order of $M_Q^2/Q^2$ and $|t|/Q^2$ have been dropped in
(\ref{ns},\ref{deld}).

The $k_{\perp}^2$ dependence  of the  $u_i$ functions, which is
important in the
calculation, has been studied. It was found that all functions
decreased
with growing $k_{\perp}^2$. A good approximation of this behaviour is
\begin{equation}
u_i(k_{\perp},r)= \frac{|t|+\mu_0}{k_{\perp}^2+|t|+\mu_0}
u_i(0,r),\;\;\;r^2=|t|,\;\;\;\mu_0\sim 1(GeV)^2. \lab{ui}
\end{equation}
This improves the convergence of the integral  over $d^2 k_{\perp}$. The
functions $u_i(0,r)$ in(\ref{ui}) are the corresponding form factors for
the on-mass-shell quarks which have been calculated perturbatively
\ci{gol4}. The result of integration over $d^2 k_{\perp}$ in (\ref{int})
is complicated in form and we do not show it here.

Expression (\ref{f2d}) has been obtained for the pomeron with
$\alpha_P(t)=1$. For the supercritical pomeron with $ \alpha_{P}(0) \ge 1$
we must replace the simple power $x_p$ by the power
$x_p^{2\alpha_{P}(t)-1}$.  This behaviour of the diffractive cross section
is connected with the pomeron flux factor (\ref{flux}).

Note that the momentum transfer $t$ was not fixed in diffractive
experiments \ci{gap1,gap2}. The integrated cross sections and
diffractive structure function
\be
F_2^{D(3)}(x,Q^2,x_p)=\int_{t_m}^{0} dt
F_2^{D(4)}(x,Q^2,x_p,t),\;\;\;|t_m|=7(GeV)^2 \lab{fd3}
\ee
are used.

Let us determine the low--$x_p$ behaviour
of  $F_2^{D(3)}$.
We find  from (\ref{f2d},\ref{nd})
\be
 F_2^{D(4)}(x,Q^2,x_p,t) \sim \frac{t e^{2bt} }{x_p^{2(\alpha_{P}
 (0)+\alpha_{P}'t)-1}}
 =\frac{t e^{2t(b+\alpha_{P}' \ln{1/x_p})}}{x_p^{2\alpha_{P}
 (0)-1}}.
\ee
Here, the exponential form of the proton form factor $F(t)=e^{bt}$   with
$b=1.9(GeV)^{-2}$ has been used.  Integration  over  $t$ in (\ref{fd3})
gives us the following form of the  $F_2^{D(3)}$ structure function at
small $x_p$
\be
 F_2^{D(3)} \propto \frac{1}{x_p^{2\alpha_{P}(0)-1}(b+\alpha_{P}'
 \ln{1/x_p})^2}.   \lab{g1x}
\ee
Using this equation we can evaluate the effective
pomeron intercept from the fit (\ref{n})
 \be
\alpha_{P}(0) =\frac{N+1}{2}+\frac{\alpha_{P}'}{b+\alpha_{P}' \ln{1/x_p}}.
\lab{resp}
\ee
We find for $x_p=10^{-4}- 10^{-2}$
 \be
\alpha_{P}(0) \sim 1.15-1.2.
\lab{respn}
\ee
This value is a little larger than the standard pomeron intercept obtained
from the elastic reactions \ci{soft} which is equal to $\alpha_{P}(0) =
1.08$.  We think that this may be connected with the manifestation of the
same pomeron in different regions of momentum transfer.  In hard scattering
processes, the interaction time is small and the single
pomeron exchange with $\alpha_{P}(0) \sim 1.2$ contributes.  In soft
diffractive hadron reactions, the interaction time is large and the
pomeron rescattering effects must be important.  These contributions
decrease the pomeron intercept to the standard value $\alpha_{P}(0) \sim
1.08$ \ci{golsup}.  The same point of view was expressed in \ci{don}.

In calculations, the perturbative results for the quark--pomeron vertex
have been used. Note that in the integral (\ref{fd3}) the integration
region over $t$ includes the range of small momentum transfer $|t| \le
1(GeV)^2$ where the nonperturbative effects should be important. However,
the contribution of this range to the structure function is rather small
because the function  $N$ in (\ref{nd}) is proportional to $|t|$. The
results of calculations for the $F_2^{D(3)}$ structure function are shown
in Fig.2 for  $\alpha_{P}(0) = 1.1$. It can be seen from this figure that
the diffractive contribution for the standard pomeron vertex is about
$15-20\%$ of the experimental results for $F_2^{D(3)}$, which coincides
with the conclusion from \ci{lanj,pom}.  The contribution of the
spin--dependent pomeron vertex increases the diffractive structure
function by a factor of about two. As we have expected from (\ref{respn}),
the experimental data have a much steeper $x_p$ dependence for this
$\alpha_{P}(0)$.

 In Fig. 3, the results of calculations  for
$\alpha_{P}(0)= 1.15$ are shown. The shape of the obtained curves coincides
with the $x_p$--dependence of experimental data. We find that the
results of theoretical calculations for the spin--dependent pomeron
vertex practically coincide in this case with  experimental results.
As previously, the predictions for the standard pomeron vertex lies
lower than the curve for the spin--dependent pomeron vertex.
Note that at small $\beta$ the invariant mass of the produced system is
large, and in addition to the effects of  Fig.1  the triple
pomeron contributions exist which should raise the
diffractive $Q\bar Q$ production effects studied here up to
experimental data (see e.g. \ci{par}).

\section{Spin-dependent Diffractive Deep
Inelastic Scattering}

Let us study now the spin--dependent cross sections
and the longitudinal double spin $A_{ll}$ asymmetry which is
determined by the relation
\be
A_{ll}=
\frac{\Delta \sigma}{\sigma}=\frac{
\sigma(^{\rightarrow} _{\Leftarrow})-\sigma(^{\rightarrow} _{\Rightarrow})}
{\sigma(^{\rightarrow} _{\Rightarrow})+\sigma(^{\rightarrow}
_{\Leftarrow})}, \lab{asydef}
\ee
where $\sigma(^{\rightarrow} _{\Rightarrow})$ and  $\sigma(^{\rightarrow}
_{\Leftarrow})$ are the cross sections with parallel and antiparallel
longitudinal polarization of lepton and proton.

The $A_{ll}$ asymmetry for the integrated cross sections over
the transverse momentum of the produced jet
have been calculated in \ci{golall}. It was found that the
asymmetry was dependent on the quark--pomeron coupling structure and could
reach $10-20\%$.

Here, we shall calculate  perturbatively the distribution of
the diffractive deep inelastic cross section
over the transverse momentum of the produced jet $k_\perp^2$.
 The difference of
the cross section for the supercritical pomeron can be written
in the form
\ba
\Delta \sigma(t)=
\frac{d^5 \sigma(^{\rightarrow} _{\Leftarrow})}{dx dy dx_p dt dk_\perp^2}-
\frac{d^5 \sigma(^{\rightarrow} _{\Rightarrow})}{dx dy dx_p dt
dk_\perp^2}= \nonumber\\
 \frac{3(2-y)\beta_0^4 F(t)^2 [9\sum_{i}e^2_i] \alpha^2}{128
x_p^{2\alpha_{P}(t)-1} Q^2 \pi^3} \frac{A(\beta,k_\perp^2,x_p,t)}
{\sqrt{1-4k_\perp^2\beta/Q^2}(k_\perp^2+M_Q^2)^2}. \lab{dsigma}
\ea
The notation here is similar to that used in Eqs. (\ref{f2d},\ref{int}).

The function $A$ is connected with the trace over the quark loop.
The leading $x_p$ dependence is extracted in the coefficient of Eq.
(\ref{dsigma}) which is determined by the pomeron flux factor. Thus,
we can calculate the function $A$ as $x_p \to 0$. It can be written as
follows:

\be
 A(\beta,k_\perp^2,t)= A^s(\beta,k_\perp^2,t)+
 \delta A(\beta,k_\perp^2,t).   \lab{na}
 \ee
Here $A^s$ is the contribution of the standard pomeron vertex
(\ref{pmu}) and $\delta A$ is determined by the
$u_1(r)-u_4(r)$ terms from (\ref{ver}).

The function $A^s$ for the light quarks looks like
\be
 A^s(\beta,k_\perp^2,t)=
   16 (2 (1-\beta) k_\perp^2 - |t| \beta) |t|   \lab{ad}
 \ee

We have calculated $\delta A$ in the $\beta \to 0$ limit.
 For the massless quarks we have
\be
 \delta A(\beta,k_\perp^2,t)= - 16 (3 k_\perp^2 + 2 |t|) k_\perp^2 |t|
    u_3.
      \lab{dad} \ee
The leading twist terms have been calculated here as in the previous
section.

The spin-average cross section can be written in the form
\ba
\sigma(t)=
\frac{d^5 \sigma(^{\rightarrow} _{\Leftarrow})}{dx dy dx_p dt dk_\perp^2}+
\frac{d^5 \sigma(^{\rightarrow} _{\Rightarrow})}{dx dy dx_p dt
dk_\perp^2}= \nonumber\\
 \frac{3(1-y+y^2/2)\beta_0^4 F(t)^2 [9\sum_{i}e^2_i] \alpha^2}{128
x_p^{2\alpha_{P}(t)} y Q^2 \pi^3} \frac{N(\beta,k_\perp^2,x_p,t)}
{\sqrt{1-4k_\perp^2\beta/Q^2}(k_\perp^2+M_Q^2)^2}, \lab{sigma}
\ea
where $N$ is determined by Eg. (\ref{nd}).

It can be seen that $\sigma$ has a more singular behavour  than $\delta
\sigma$ as $x_p \to 0$. This is determined by the fact that the leading
term in $\delta \sigma$ is proportional to
$\epsilon^{\mu\nu\alpha\beta}r_\beta...\propto x_p p$. Similar is true for
the lepton part of the diagram of Fig.1. As a result, the
additional term $y x_p$ appears in $\delta \sigma$.

 As in the previous section, we shall calculate the cross section
 integrated over momentum transfer
\be
\sigma[\Delta \sigma]=\int_{t_m}^{0} dt
\sigma(t)[\Delta \sigma(t)],\;\;\;|t_m|=7(GeV)^2. \lab{intsi}
\ee

The expressions for the standard pomeron coupling contributions to $\sigma$
and $\Delta \sigma$ have been found for arbitrary $\beta$. However, the
results for the spin--dependent part of the pomeron coupling have been
obtained for $\beta=0$.  So, we shall calculate the cross sections and
asymmetry for not very large $\beta=0.175$. The results of calculation for
the cross section of the light quark production in diffractive deep
inelastic scattering for the pomeron with the usual intercept
$\alpha_{P}(0)= 1.1$ and $y=0.7$ are shown in Fig. 4 for the standard and
spin-dependent pomeron couplings.  The predicted cross sections are not
small. The shape of  both the curves is very similar and for the
spin-dependent pomeron coupling the cross section is larger by a factor of
about two, as previously.

The asymmetry in the diffractive $Q \bar Q$ production is shown in Fig. 5.
It can be seen from the cross section (\ref{dsigma},\ref{sigma}) that
the asymmetry for the standard quark--pomeron vertex is very simple in
form
\be
A_{ll}=\frac{y x_p (2-y)}{2-2y+y^2}.
\ee
There is no any $k_\perp$ and $\beta$ dependence here. For the
spin--dependent pomeron coupling the asymmetry is more  complicated
because of the different contributions in $\delta A$ and
$\delta N$  proportional to $k^2_{\perp}$.
In this case the $A_{ll}$ asymmetry is smaller than for the standard
 pomeron vertex. Thus, one can use the
$A_{ll}$ asymmetry to test the quark-pomeron coupling structure.  Note
 that from $\Delta \sigma$  in (\ref{dsigma}) the diffractive contribution
to the $g_1$ spin--dependent structure function can be determined \ci{g1g}.
The obtained low -$x$ behaviour of $ g_1(x)$ has a singular form like
$1/(x^{0.3} \ln^2(x))$ which is compatible with the SMC data for
$g^p_1(x)$ \ci{smc}.

 \section{Conclusion}

 Thus, we have found that the  structure of the quark--pomeron coupling can
 affect spin average and spin--dependent cross section.
From the analysis of the $F_2^D$ structure function we have found that the
slope of the HERA experimental data at small $x_p$ leads to the pomeron
intercept of about $\alpha_P(0) \sim 1.15$, which is a little larger than
$\alpha_P(0) \sim 1.08$ for elastic reactions.

The spin--dependent form of $V_{qqP}$ leads to increasing of the cross
section by a factor of about two. However, the shape of the cross sections
 is very similar for the standard and spin--dependent pomeron vertices.
As a result, it will be difficult to test the pomeron coupling structure
from the analysis of the cross section. The $A_{ll}$ asymmetry is more
convenient for this purpose. The asymmetry is free from all normalization
factors and sensitive to the dynamics of pomeron interaction.  Moreover,
we have a well-defined prediction for $A_{ll}$ for the standard pomeron
vertex. This conclusion is similar to the results of \ci{goltr} where
the single-spin asymmetry in the diffractive $Q \bar Q$ production has been
studied.

It has been mentioned that at small $\beta$
there is a contribution which is associated with the triple pomeron
interaction. The relative  role of the diffractive $Q \bar Q$  production
and $PPP$ contribution  can be tested experimentally by the observation of
two high--$p_t$ jet events and events that have more that two jets.

So, in this paper the perturbative QCD analysis of the effects determined
by the spin--dependent quark-pomeron coupling in the diffractive deep
inelastic scattering has been done. The
sensitivity  of the spin  dependent cross section to the quark--pomeron
coupling structure has been found. The
nonperturbative contribution  might be important in these reactions too.

We can conclude that the investigation of the longitudinal double
spin asymmetry and the cross section of the diffractive deep inelastic
scattering can give important information about the complicated spin
structure or the pomeron coupling. For testing the pomeron--proton
vertex the transverse polarization of the proton target (beam) should be
more relevant \ci{goltr}. The HERA facilities to study properties of
the pomeron can give a possibility to test the size of the spin--flip
pomeron coupling.\\[1cm]

 The author expresses his deep gratitude to  N.Akchurin, A.V.Efremov,
 G.Mallot, P.Kroll, W.-D.Nowak, A.Penzo, G.Ramsey, A.Sch\"afer,
O.V.Teryaev for fruitful discussions.

\newpage

{\bf Figure captions}\\
{\bf Fig.1} ~Diffractive $Q \bar Q$   production in deep
inelastic scattering.\\[.5cm]
{\bf Fig.2} ~ $Q \bar Q$   production contribution to the $F_2^{D(3)}$
diffractive structure function at small $x_p$.  Theoretical curves are
shown for $Q^2=25(GeV)^2$ and $\alpha_{P}(0)= 1.1$:  solid line -for the
standard quark-pomeron vertex; dot-dashed line -for the spin-dependent
vertex. Data are from Ref.  \ci{gap2}.\\[.5cm]
{\bf Fig.3}  $Q \bar Q$   production contribution to the ~$F_2^{D(3)}$
diffractive structure function at small $x_p$.  Theoretical curves are
shown for $Q^2=25(GeV)^2$ and $\alpha_{P}(0)= 1.15$:  solid line -for the
standard quark-pomeron vertex; dot-dashed line -for the spin-dependent
vertex. Data are from Ref.  \ci{gap2}.\\[.5cm]
{\bf Fig.4} ~Distribution of $\sigma$ over jets $k^2_{\perp}$.
Solid line -for the standard vertex;
dot-dashed line -for the spin-dependent quark-pomeron vertex.\\[.5cm]
{\bf Fig.5} ~$k^2_{\perp}$-- dependence of $A_{ll}$ asymmetry.
Solid line -for the standard vertex;
dot-dashed line -for the spin-dependent quark-pomeron vertex.

\newpage

  \vspace*{.5cm}
 \hspace{4.5cm}
\epsfxsize=12cm
{\epsfbox{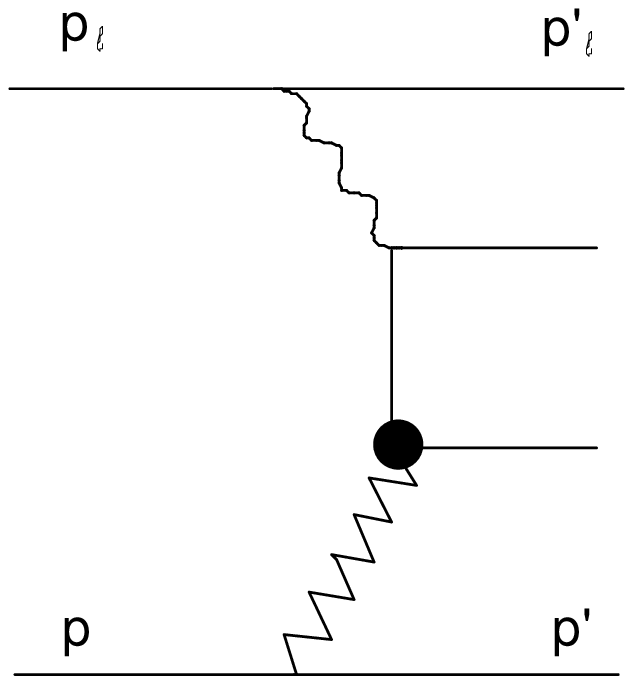}}
  \vspace*{-11.6cm}
\begin{center}
Fig.1
\end{center}

\vspace{2cm}
\samepage
  \vspace*{-.1cm}
\epsfxsize=13cm
\centerline{\epsfbox{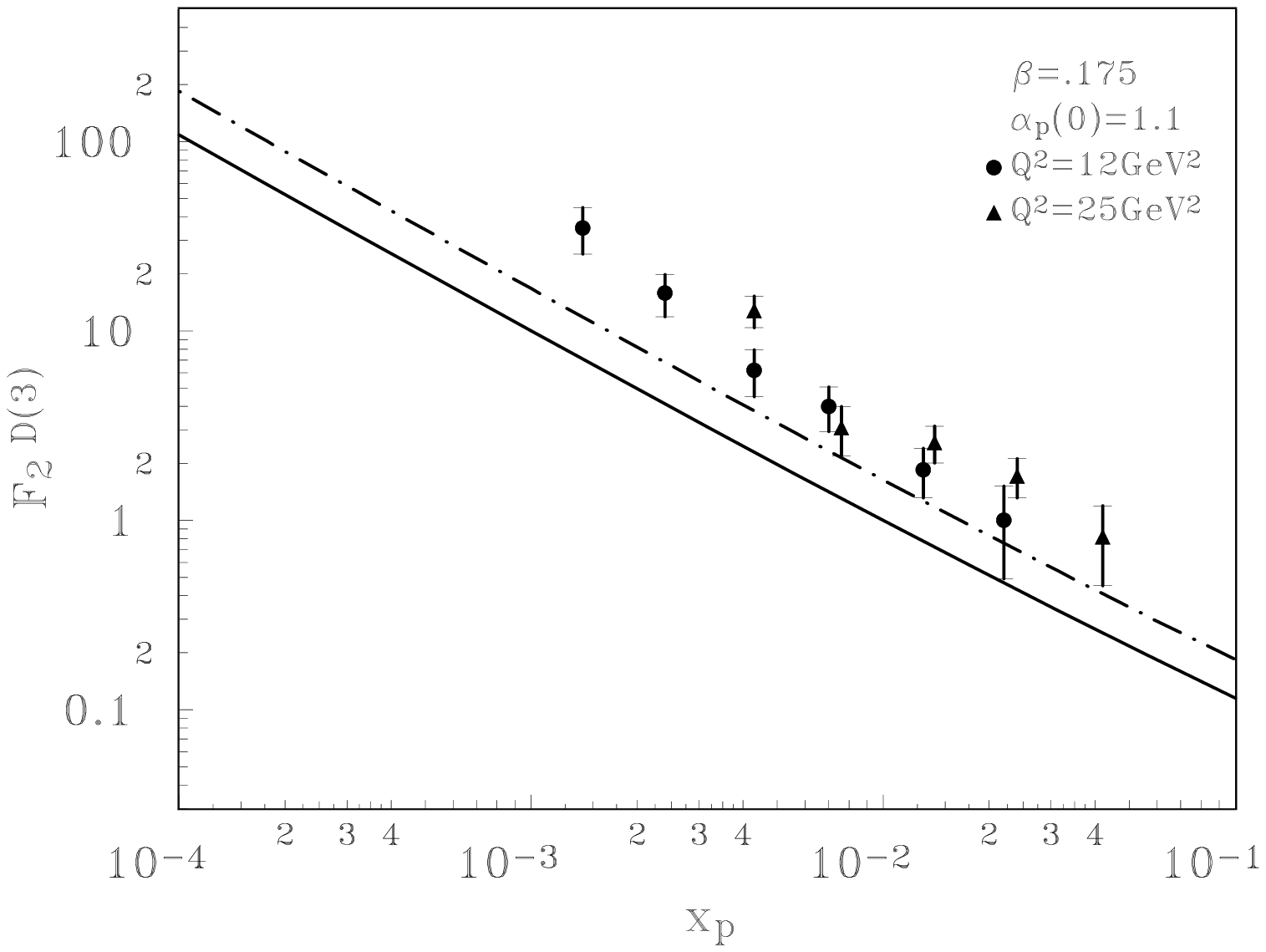}}
  \vspace*{.3cm}
\begin{center}
Fig.2
\end{center}

\newpage
  \vspace*{-.5cm}
\epsfxsize=13cm
\centerline{\epsfbox{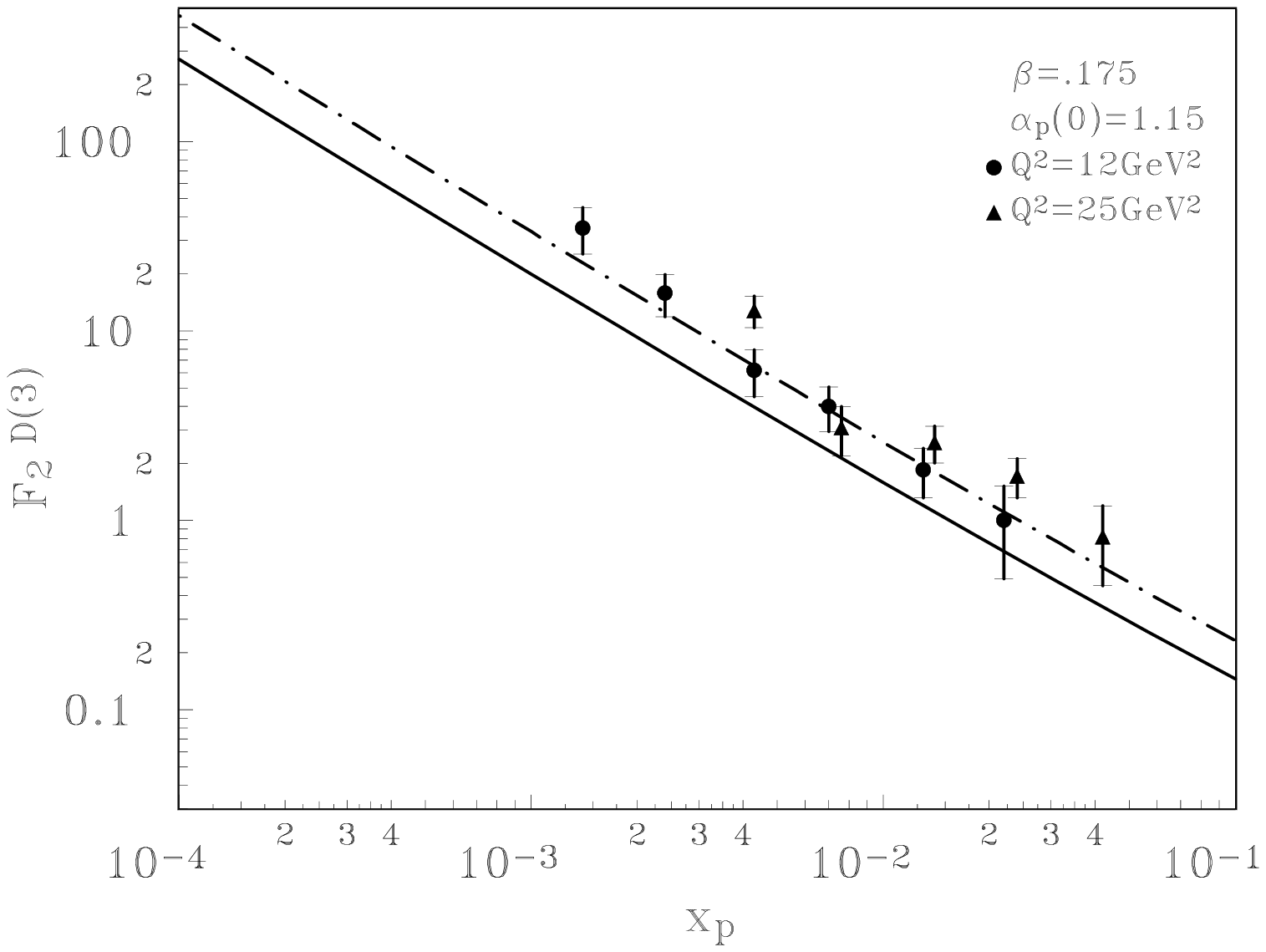}}
  \vspace*{.2cm}
\begin{center}
Fig.3
\end{center}

  \vspace*{-.1cm}
       \hspace*{.5cm}
\epsfxsize=13cm
{\epsfbox{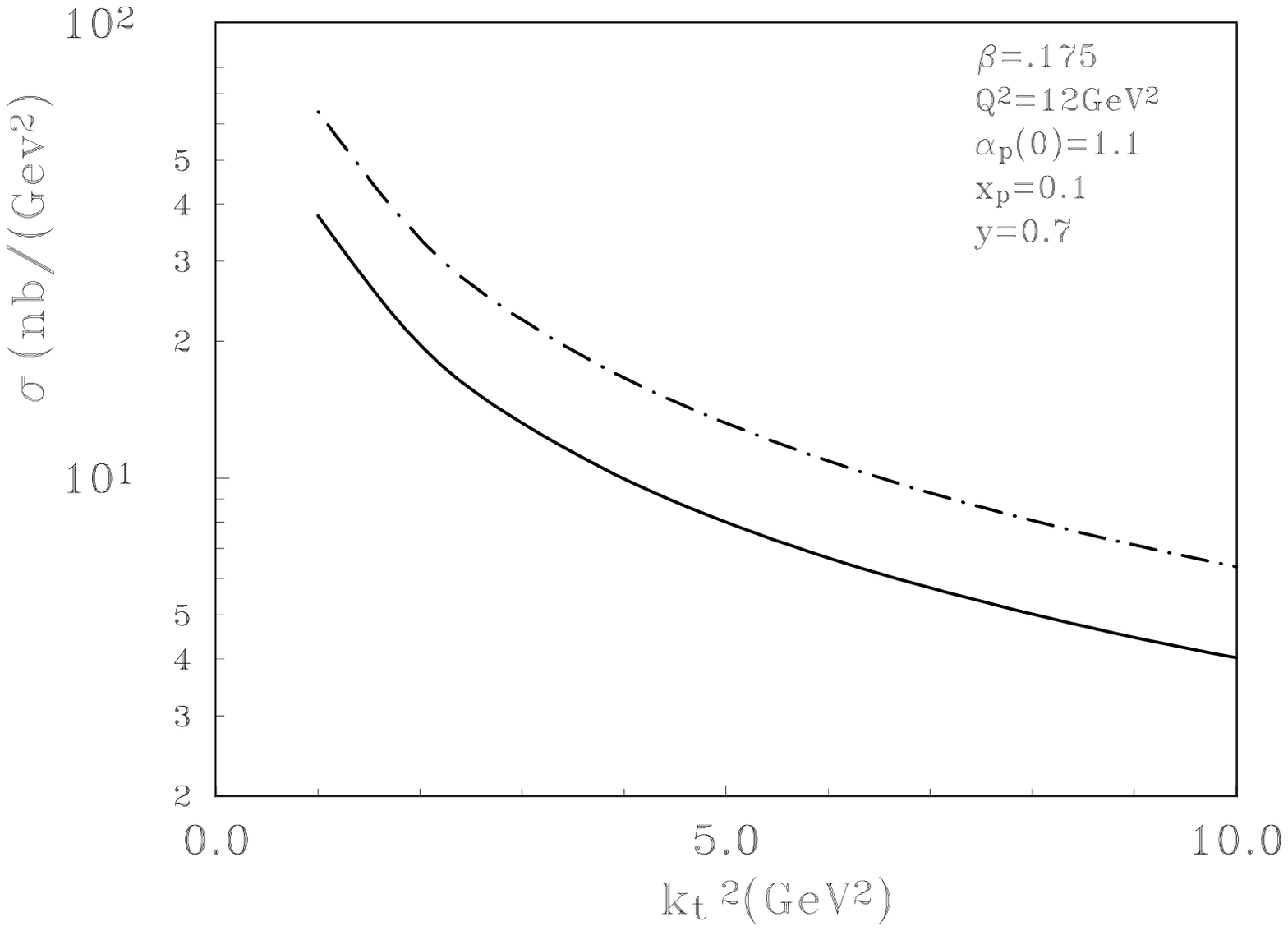}}
  \vspace*{.3cm}
\begin{center}
Fig.5
\end{center}

\newpage
  \vspace*{.5cm}
         \hspace*{.5cm}
       \epsfxsize=13cm
{\epsfbox{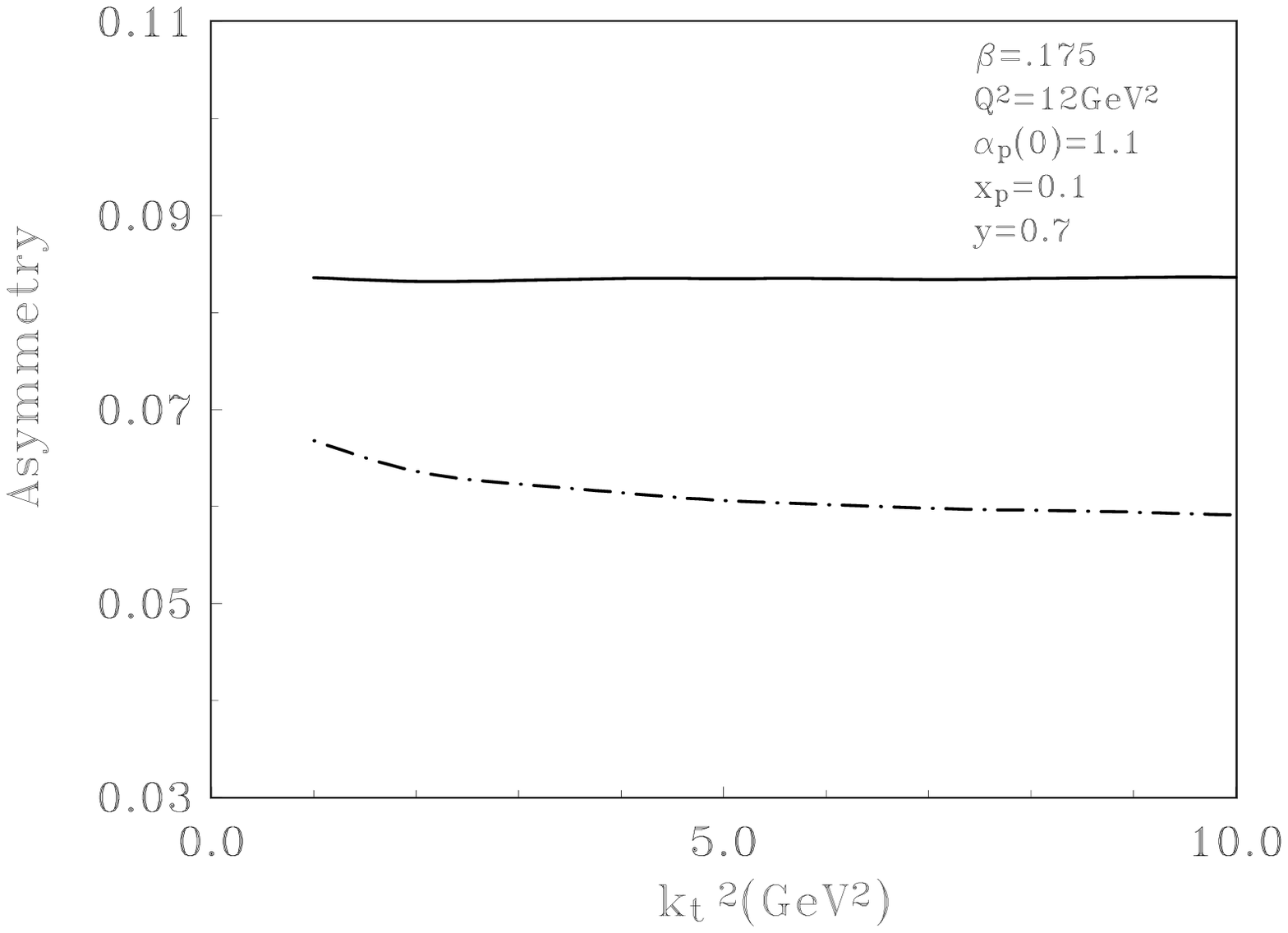}}
  \vspace*{.2cm}
\begin{center}
Fig.5
\end{center}

\end{document}